\documentclass[aps,pra,superscriptaddress,showpacs,amsmath,amssymb]{revtex4-1}
\usepackage{graphicx}

\newcommand*{\diag}{\mathop{\mathrm{diag}}}

\begin{document}

\title{Transfer of orbital angular momentum of light using two component
slow light}

\author{Julius Ruseckas}
\email{julius.ruseckas@tfai.vu.lt}
\homepage{http://www.itpa.lt/~ruseckas}
\affiliation{Institute of Theoretical Physics and Astronomy, Vilnius University,
A. Go\v{s}tauto 12, Vilnius 01108, Lithuania}

\author{Via\v{c}eslav Kudria\v{s}ov}
\affiliation{Institute of Theoretical Physics and Astronomy, Vilnius University,
A. Go\v{s}tauto 12, Vilnius 01108, Lithuania}

\author{Ite A. Yu}
\affiliation{Department of Physics and Frontier Research Center on Fundamental
and Applied Sciences of Matters, National Tsing Hua University, Hsinchu
30013, Taiwan}

\author{Gediminas Juzeli\=unas}
\email{gediminas.juzeliunas@tfai.vu.lt}
\affiliation{Institute of Theoretical Physics and Astronomy, Vilnius University,
A. Go\v{s}tauto 12, Vilnius 01108, Lithuania}

\date{\today}

\begin{abstract}
We study the manipulation of slow light with an orbital angular momentum
propagating in a cloud of cold atoms. Atoms are affected by four copropagating
control laser beams in a double tripod configuration of the atomic energy levels
involved, allowing to minimize the losses at the vortex core of the control
beams. In such a situation the atomic medium is transparent for a pair of
copropagating probe fields, leading to the creation of two-component (spinor)
slow light. We study the interaction between the probe fields when two control
beams carry optical vortices of opposite helicity. As a result, a transfer of
the optical vortex takes place from the control to the probe fields without
switching off and on the control beams. This feature is missing in a single
tripod scheme where the optical vortex can be transferred from the control to
the probe field only during either the storage or retrieval of light.
\end{abstract}


\maketitle

\section{Introduction}

Distinctive properties of slow \cite{Hau-99,Kash99,Budker99,Novikova07PRL,Davidson09NP},
stored \cite{Fleischhauer-00,Liu-01,Phill2001,Juzeliunas-02,Scully02PRL,Yanik05PRA,Gor05EPJD,Lukin05Nature,Ite07PRA,Hau07Nature,Bloch09PRL,Hau09PRL,Beil10PRA}
and stationary \cite{Stationary-light-03--09,Stationary-light2,Stationary-light4,Stationary-ligth3,Stationary-ligth5,Otterbach10PRL}
light have been extensively studied for more than a decade. The research
has been motivated both by the fundamental interest in the slow and
stationary light and also because of the potential applications including
inter alia the reversible quantum memories \cite{Fleischhauer-00,Juzeliunas-02,Scully02PRL,Lukin05Nature,Lukin-03,Fleischhauer-05,Appel08PRL,Honda08PRL,Akiba09NJP}
and non-linear optics at low intensities \cite{Imamoglu96OL,Imamoglu98OL,Lukin11PRL,Lukin12Nature,Yu12PRL}.
The slow light is formed due to the phenomenon known as the electromagnetically
induced transparency (EIT) \cite{Arimondo-96,Harris-97,Scully-97,Lukin-03,Fleischhauer-05}.
The EIT emerges in a medium resonantly driven by several laser fields
and involves the destructive quantum interference between different
resonant excitation pathways of atoms. As a result a weaker (probe)
beam of light tuned to an atomic resonance can propagate slowly and
almost lossles when the medium is driven by one or several control
beams light with a higher intensity. Under suitable conditions reshaping
of the dispersive properties of the medium by the control beams leads
to a drastic reduction in the group velocity of the probe pulse. Group
velocities as small as several of tens of meters per second have been
reported \cite{Hau-99,Kash99,Budker99,Novikova07PRL,Davidson09NP}.
Most of the work on the slow light dealt with a single probe beam
and one or several control beams resonantly interacting with atomic
media, with $\Lambda$ configuration of the atom-light coupling being
one the mostly exploited \cite{Arimondo-96,Harris-97,Scully-97,Lukin-03,Fleischhauer-05}. 

Recently it was suggested to create a two component slow light using
a more complex double tripod setup \cite{Unanyan-PRL-2010,Ruseckas-PRA-2011b}
involving three ground atomic states coupled with two excited states.
Such a setup supports a simultaneous propagation of two probe beams
and leads to the formation of a two-component slow light. By properly
choosing the control lasers one can generate a tunable coupling between
the constituent probe fields. 

The orbital angular momentum (OAM) \cite{Allen-99,Allen-03,Andrews2008OAM-Book,Padget11AOP}
provides an additional possibility in manipulating the slow light.
The optical OAM represents an extra degree of freedom which can be
exploited in the quantum computation and quantum information storage
\cite{Padget11AOP}. Most of the previous studies on the vortex slow
light considered situations where the incident probe beam carries
an OAM \cite{Ruost04PRL,Davidson07PRL,Ruseckas-07,Yelin08PRA,Moretti09PRA},
yet the control beam has no vortex. Application of a vortex control
beam causes a potential problem, because its intensity goes to zero
at the vortex core leading to the disappearance of the EIT accompanied
with the absorption losses in this spatial region. To avoid such
losses it was suggested \cite{Ruseckas-PRA-2011a} to employ an extra
control laser beam without an optical vortex making a more complex
tripod scheme of the atom-light coupling previously considered for
the non-vortex beams of light \cite{Unanian,Knight-02,Rebic04PRA,Petrosyan04PRA,Yelin04PRA,Rus05,Mazets05PRA,Zaremba06OC,Zaremba07PRA,Zaremba07OC,Garva07OC}.
The total intensity of the control lasers is then non-zero at the
vortex core of the first control laser thus avoiding the losses. Using
such a scheme a transfer of an optical vortex can be accomplished
during the switching off and on the control beams \cite{Ruseckas-PRA-2011a}.

Here we show that the transfer of the vortex between the control and
probe beams can be accomplished without switching off and on the
control beams using a more complex double tripod scheme of the atom-light
coupling. The scheme shown in the Fig.~\ref{fig:double-tripod} involves
three atomic ground states coupled to two excited states via four
control beams two of them carrying optical vortices. If the incoming
probe beam does not carry an optical vortices, the coupling with
the control beams generates another component of the probe beam containing
an OAM. We explore the efficiency of such a transfer of the
optical vortex. We analyze the losses resulting from the exchange
of the optical vortex between the control and probe beams, and provide
conditions for the optical vortex of the control beam to be transferred
efficiently to the second component of the probe beam. 

The paper is organized as follows. In Sec.~\ref{sec:formulation}
we present the double tripod scheme and the equations for atomic operators
and probe fields. In Sec.~\ref{sec:propagation} we derive equations
of propagation for the probe fields using an adiabatic approximation.
We use those equations in Sec.~\ref{sec:vortex} for the description
of the transfer of OAM from control to probe beams. Section \ref{sec:conclusions}
summarizes the findings.

\section{\label{sec:formulation}Formulation}

\begin{figure}
\includegraphics[width=0.33\textwidth]{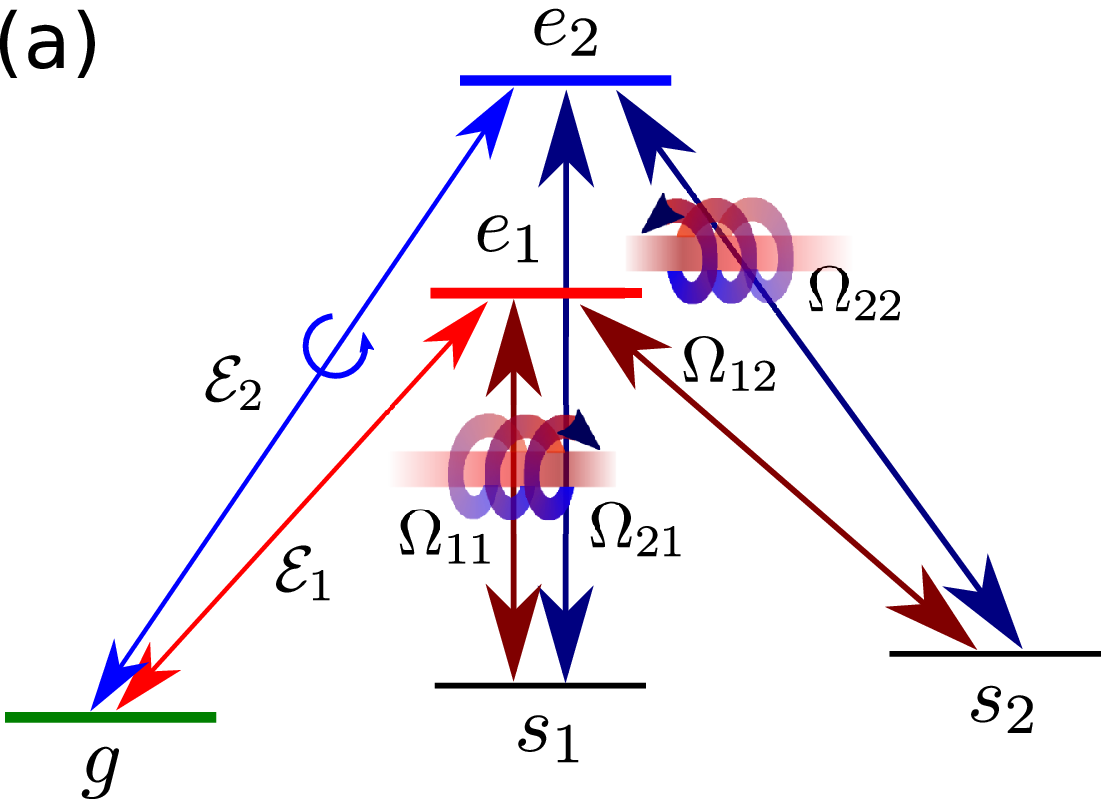}\includegraphics[width=0.4\textwidth]{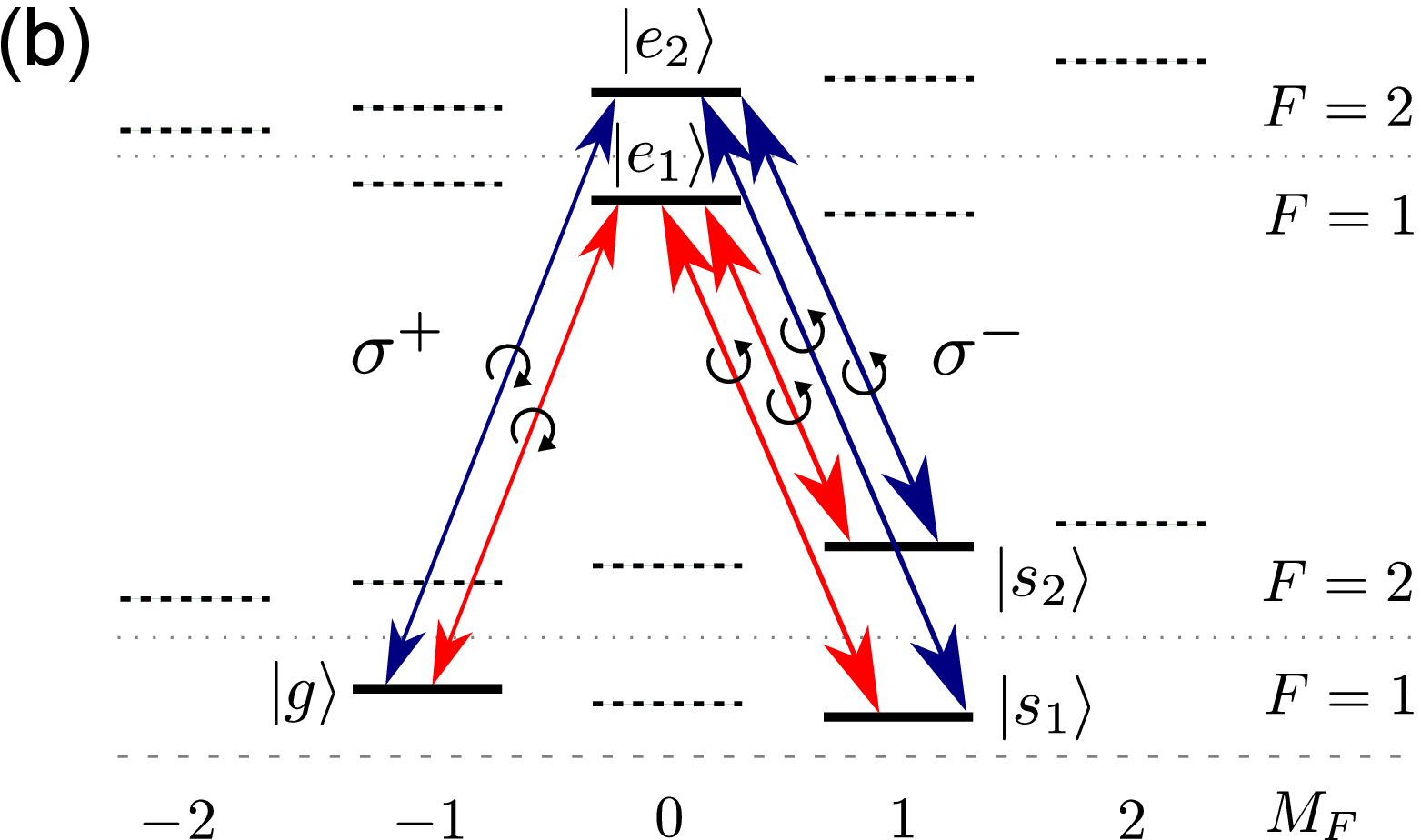}
\caption{(Color online) (a) Double tripod level scheme.
(b) Possible experimental realization of the double tripod setup for atoms
like Rubidium \cite{Phillips-PRL-2001} or Sodium \cite{Liu-Nature-2001}.
The scheme involves transitions between the magnetic states of two
hyperfine levels with $F=1$ and $F=2$ for the ground and excited state
manifolds. Both probe beams are circular $\sigma^{+}$ polarized and all
four control beams are circular $\sigma^{-}$ polarized.}
\label{fig:double-tripod}
\end{figure}

We shall consider the light-matter interaction in an ensemble of atoms
using a double tripod coupling scheme shown in Fig.~\ref{fig:double-tripod}(a).
The atoms are characterized by three hyperfine ground levels $|g\rangle$,
$|s_{1}\rangle$ and $|s_{2}\rangle$ which have dipole-allowed optical
transitions to the electronic excited state levels $|e_{1}\rangle$
and $|e_{2}\rangle$. The atom-light coupling scheme involves
two laser fields of low intensity (probe fields) and four fields of
much higher intensity (control fields). The probe beams are described
by the electric field amplitudes $\mathcal{E}_{1}$ and $\mathcal{E}_{2}$
with the corresponding central frequencies $\omega_{1}$ and $\omega_{2}$.
They drive atomic transitions $|g\rangle\rightarrow|e_{1}\rangle$
and $|g\rangle\rightarrow|e_{2}\rangle$, characterized by the dipole
moments $\mu_{1}$ and $\mu_{2}$. Control fields having frequencies
$\omega_{jq}$ couple the atomic transitions $|s_{1}\rangle\rightarrow|e_{1}\rangle$
and $|s_{2}\rangle\rightarrow|e_{2}\rangle$ with coupling strength
being characterized by Rabi frequencies $\Omega_{jq}$, $(j,q=1,2)$
. We assume the four photon resonances with the probe beams for each
pair of the control lasers, meaning that $\omega_{1}-\omega_{1q}=\omega_{2}-\omega_{2q}$.
The presence of the control beams makes the medium transparent for the resonant probe
beams in a narrow frequency range due to the electromagnetically induced
transparency (EIT). To satisfy the EIT conditions probe fields
should be quasimonochromatic requiring that the amplitudes $\mathcal{E}_{j}$
change a little during the optical cycle.

The double tripod scheme can be implemented using atoms like Rubidium or Sodium
which contain two hyperfine ground levels with $F=1$ and $F=2$, as illustrated
in Fig.~\ref{fig:double-tripod}(b). Such atoms have been used in the initial
light-storage experiments involving a simpler $\Lambda$ setup
\cite{Liu-Nature-2001,Phillips-PRL-2001}. In the present context the states
$|g\rangle$ and $|s_1\rangle$ correspond to the magnetic states with $M_F=-1$
and $M_F=1$ of the $F=1$ hyperfine ground level, whereas the state
$|s_2\rangle$ represents the hyperfine ground state with $F=2$ and $M_F=1$. The
two states $|e_1\rangle$ and $|e_2\rangle$ correspond to the electronic excited
states with $M_F=0$ of the $F=1$ and $F=2$ manifolds. To make the double tripod
setup both probe beams are to be circular $\sigma^{+}$ polarized and all four
control beams are to be circular $\sigma^{-}$ polarized. Note that such a
scheme can be implemented by adding three extra control laser beams to the
$\Lambda$ setup used previously in the experiment by Liu \textit{et al}
\cite{Liu-Nature-2001}.

In the previous studies of the multicomponent light \cite{Unanyan-PRL-2010,Ruseckas-PRA-2011b}
the counterpropagating probe and control beams have been considered.
Here we analyze an opposite situation where the probe and control
fields co-propagate (along the $z$ axis). Probe fields can then be
written as $\mathcal{E}_{1}(\mathbf{r},t)e^{ik_{1}z}$ ,
$\mathcal{E}_{2}(\mathbf{r},t)e^{ik_{2}z}$,
with $k_{j}=\omega_{j}/c$ being the central wave-vector of the $j$-th
probe beam. The copropagating setup is more suited for the efficient
transfer of an optical vortex between the control and probe beams
we are interested in. For paraxial beams the amplitudes $\mathcal{E}_{1}(\mathbf{r},t)$
and $\mathcal{E}_{2}(\mathbf{r},t)$ depend weakly on the propagation
direction $z$, the fast spatial dependence being accommodated in
the exponential factors $e^{ik_{j}z}$. The same applies to the control
beams which copropagate along the $z$ axis and has the form $\Omega_{jq}e^{ik_{jq}z}$,
with $k_{jq}=\omega_{jq}/c$ being the central wave-vector. 

We shall neglect the atomic center of mass motion. The electronic
properties of the atomic ensemble is described by the atomic flip
operators $\sigma_{e_{j}g}$ and $\sigma_{s_{j}g}$ describing the
coherences between the atomic internal states $|e_{j}\rangle$ , $|s_{j}\rangle$
and $|g\rangle$ at a certain spatial point. For the convenience
we introduce the column of the probe field amplitudes $\mathcal{E}=(\mathcal{E}_{1},\mathcal{E}_{2})^{T}$
and columns of atomic flip operator $\sigma_{eg}=(\sigma_{e_{1}g},\sigma_{e_{2}g})^{T}$
and $\sigma_{sg}=(\sigma_{s_{1}g},\sigma_{s_{2}g})^{T}$. Defining
the parameter $g=g_{j}=\mu_{j}(\omega_{j}/2\varepsilon_{0}\hbar)^{1/2}$
characterizing the atom-light coupling strength (assumed to be equal
for both probe fields), the equation for the slowly in time and in
space varying amplitudes of the probe fields can be written as follows:
\begin{equation}
\partial_{t}\mathcal{E}+c\partial_{z}\mathcal{E}-i\frac{1}{2}c\hat{k}^{-1}\nabla_{\bot}^{2}\mathcal{E}=ign\sigma_{eg}\,,\label{eq:s-electric-0}
\end{equation}
where $n$ denotes the atomic density and $\hat{k}=\diag(k_{1},k_{2})$
is a diagonal $2\times2$ matrix of the probe field wavevectors.

The diffraction term containing the transverse derivatives
$\nabla_{\bot}^{2}\mathcal{E}$ can be neglected when the change in the phase of
the probe fields due to this term is much smaller than $\pi$. The transverse
derivative can be estimated as
$\nabla_{\bot}^{2}\mathcal{E}\sim\sigma^{-2}\mathcal{E}$, where $\sigma$ is a
characteristic transverse dimension of the probe beams.  If the probe beam
carries an optical vortex, $\sigma$ can be associated with a width of the
vortex core. On the other hand, for the probe beam without an optical vortex,
$\sigma$ is a characteristic width of the beam.  The change in time of the
probe field can be estimated as $\partial_{t}\mathcal{E}\sim
cL^{-1}\Delta\mathcal{E}$, where $L$ is the length of the atomic cloud and
$\Delta\mathcal{E}$ is the change of the field. Thus the change of the phase
due to the diffraction term is $L/2k\sigma^{2}$. It can be neglected when the
sample length $L$ is not too large, $L\lambda/\sigma^{2}\ll1$. Taking the
length of the atomic cloud $L=100\,\mu\mathrm{m}$, the characteristic
transverse dimension of the probe beam $\sigma=20\,\mu\mathrm{m}$ and the
wave-length $\lambda=1\,\mu\mathrm{m}$, we obtain $L\lambda/\sigma^{2}=0.25$.
Therefore we can drop out the diffraction term in Eq.~(\ref{eq:s-electric-0})
obtaining
\begin{equation}
\partial_{t}\mathcal{E}+c\partial_{z}\mathcal{E}=ign\sigma_{eg}\,.\label{eq:s-electric}
\end{equation}
Equations describing the atom-light coupling are
\begin{eqnarray}
i\partial_{t}\sigma_{eg} & = & -i\gamma\sigma_{eg}-\hat{\Omega}\sigma_{sg}-g\mathcal{E}\,,\label{eq:e}\\
i\partial_{t}\sigma_{sg} & = & \hat{\delta}\sigma_{sg}-\hat{\Omega}^{\dag}\sigma_{eg}\,,\label{eq:s}
\end{eqnarray}
where the dagger refers to a Hermitian conjugated matrix. The equations
are treated in the frames of reference rotating with frequencies $\omega_{j}$
and $\omega_{j}-\omega_{jj}$, respectively. Here
$\hat{\delta}=\diag(\delta_{1},\delta_{2})$
is a diagonal $2\times2$ matrix of two photon detunings with $\delta_{q}=\omega_{s_{q}g}+\omega_{1q}-\omega_{1}=\omega_{s_{q}g}+\omega_{2q}-\omega_{2}$,
$\hat{\Omega}$ is a $2\times2$ matrix of Rabi frequencies with matrix
elements $\Omega_{ij}$ and $\gamma$ is the decay rate of excited
levels. The decay rate $\gamma$ is assumed to be the same for both
levels $|e_{1}\rangle$ and $|e_{2}\rangle$. Initially the atoms
are in the ground level $g$ and the Rabi frequencies of the probe
fields are considered to be much smaller than those of the control
fields. Consequently one can neglect the depletion of the ground level
$|g\rangle$.

\section{\label{sec:propagation}Propagation of the probe beams}

Equations (\ref{eq:e})--(\ref{eq:s}) provide two limiting cases.
If the Rabi frequencies of the control beams driving transitions from
the level $|s_{2}\rangle$ are proportional to the Rabi frequencies
of the beams driving transitions from the level $|s_{1}\rangle$ ($\Omega_{22}/\Omega_{21}=\Omega_{12}/\Omega_{11}$),
the double tripod system becomes equivalent to a double $\Lambda$
system for zero two photon detuning. On the other hand if $\Omega_{11}\Omega_{21}^{*}+\Omega_{12}\Omega_{22}^{*}=0$,
the double tripod system is equivalent to two not connected tripod
systems for zero two photon detuning. We shall concentrate on the
case where the double tripod system is not equivalent to a double
$\Lambda$ system and the inverse matrix $(\hat{\Omega}^{\dag})^{-1}$
does exist. Thus Eq.~(\ref{eq:s}) relates $\sigma_{eg}$ to $\sigma_{sg}$
as:
\begin{equation}
\sigma_{eg}=(\hat{\Omega}^{\dag})^{-1}(\hat{\delta}-i\partial_{t})\sigma_{sg}\,.\label{eq:e-s}
\end{equation}

\subsection{Adiabatic approximation}

In what follows the control and probe beams are considered to be close
to the two-photon resonance. Application of such resonant beams cause
the electromagnetically induced transparency (EIT) in which the optical
transitions from the atomic ground states $|g\rangle$, $|s_{1}\rangle$
and $|s_{2}\rangle$ interfere destructively preventing population
of the excited states $|e_{1}\rangle$ and $|e_{2}\rangle$. The adiabatic
approximation is obtained neglecting the population of the latter
excited states described by the spin-flip operator $\sigma_{eg}$
in Eq.~(\ref{eq:e}). Thus one has 
\begin{equation}
\sigma_{sg}=-g\hat{\Omega}^{-1}\mathcal{E}\,.\label{eq:s-g}
\end{equation}
Equations (\ref{eq:s-electric}), (\ref{eq:e-s}), and (\ref{eq:s-g})
provide a closed set of equations for the electric field amplitudes
$\mathcal{E}_{1}$ and $\mathcal{E}_{2}$. Assuming the control beams
to be time-independent, one arrives at the following matrix equation
for the column of the probe fields
\begin{equation}
(c^{-1}+\hat{v}^{-1})\partial_{t}\mathcal{E}+\partial_{z}\mathcal{E}+i\hat{v}^{-1}\hat{D}\mathcal{E}=0\,,\label{eq:result1}
\end{equation}
where 
\begin{equation}
\hat{D}=\hat{\Omega}\hat{\delta}\hat{\Omega}^{-1}
\end{equation}
is a matrix of the two-photon detuning and
\begin{equation}
\hat{v}=\frac{c}{g^{2}n}\hat{\Omega}\hat{\Omega}^{\dag}\label{eq:gr-vel}
\end{equation}
is the matrix of group velocity. If the two-photon detunings $\delta_{1}$
and $\delta_{2}$ are zero ($\hat{D}=0$), the last term drops out
in the equation of motion (\ref{eq:result1}). 

For generality the group velocity matrix $\hat{v}$ is not diagonal
and thus the probe fields $\mathcal{E}_{1}$ and $\mathcal{E}_{2}$
do not have a definite group velocity. This leads to the mixing between
fields $\mathcal{E}_{1}$ and $\mathcal{E}_{2}$. We shall return
to this issue in the following Section.

\subsection{Non-adiabatic corrections}

In order to obtain non-adiabatic corrections one needs to include
the decay rate $\gamma$ of the exited levels. Substituting Eq.~(\ref{eq:e-s})
into Eq.~(\ref{eq:e}) yields
\begin{equation}
\sigma_{sg}=-g\hat{\Omega}^{-1}\mathcal{E}-i\gamma\hat{\Omega}^{-1}(\hat{\Omega}^{\dag})^{-1}(\hat{\delta}-i\partial_{t})\sigma_{sg}\,,\label{eq:s-2}
\end{equation}
where the decay rate $\gamma$ is assumed to be much larger than the
rate of changes of $\sigma_{eg}$. Equation (\ref{eq:s-2}) can be
solved iteratively. Equation (\ref{eq:s-g}) is the first-order solution.
Substituting expression (\ref{eq:s-g}) for $\sigma_{sg}$ into
Eq.~(\ref{eq:s-2}) one arrives at the second-order solution
\begin{equation}
\sigma_{sg}=-g\hat{\Omega}^{-1}\mathcal{E}+i\gamma g\hat{\Omega}^{-1}(\hat{\Omega}^{\dag})^{-1}(\hat{\delta}-i\partial_{t})\hat{\Omega}^{-1}\mathcal{E}\,.
\end{equation}
This leads to a more general equation for the propagation of the
probe fields (EIT polaritons) in the atomic cloud
\begin{equation}
(c^{-1}+\hat{v}^{-1})\partial_{t}\mathcal{E}+\partial_{z}\mathcal{E}+i\hat{v}^{-1}\hat{D}\mathcal{E}+\frac{c\gamma}{g^{2}n}[\hat{v}^{-1}(\hat{D}-i\partial_{t})]^{2}\mathcal{E}=0\,.\label{eq:result-2}
\end{equation}
The last term represents non-adiabatic correction providing a finite
life-time for the polaritons.

\section{\label{sec:vortex}Transfer of an optical vortex}

\subsection{Control beams with optical vortices}

Up to now no assumption has been made concerning the spatial profile
of the control beams. In the following the control beams with Rabi
frequencies $\Omega_{11}$ and $\Omega_{22}$ are assumed to carry
optical vortices. Specifically, we take the intensities to be equal
$|\Omega_{11}|=|\Omega_{22}|$ and vorticities to be opposite: $l_{11}=-l_{22}\equiv l$.
Another two non-vortex control beams also have equal amplitudes, $|\Omega_{12}|=|\Omega_{21}|$,
yet there might be a phase difference $2S$ between the fields. Under
these conditions the amplitudes of the control beams can be written
as
\begin{eqnarray}
\Omega_{11} & = & |\Omega_{11}|e^{il\varphi}\:,\qquad\Omega_{22}=|\Omega_{11}|e^{-il\varphi}\label{eq:Rabi-1}\\
\Omega_{12} & = & |\Omega_{12}|\:,\qquad\Omega_{21}=|\Omega_{12}|e^{-i2S}\label{eq:Rabi-2}
\end{eqnarray}
When $S=\pi/2$ and two-photon detunings are zero, two independent tripods are formed
all over the space. Furthermore if $|\Omega_{11}|=0$, two independent tripods are formed
for any value of $S$. This takes place at the axis of the optical
vortex. On the other hand, if $S=0$ and $|\Omega_{11}|=|\Omega_{12}|\ne0$,
one arrives at the double-lambda case for which the inverse velocity
matrix becomes singular.

Introducing the angle
\begin{equation}
\tan\phi=\frac{|\Omega_{11}|}{|\Omega_{12}|}
\end{equation}
and the total Rabi frequency
\begin{equation}
\Omega(\rho)=\sqrt{|\Omega_{12}|^{2}+|\Omega_{11}|^{2}}\,,\label{eq:Omega}
\end{equation}
the eigenvalues of the velocity matrix have the form
\begin{equation}
v^{\pm}=v_{0}(1\pm\cos(S)\sin(2\phi))\,,
\end{equation}
with
\begin{equation}
v_{0}(\rho)=\frac{c\Omega^{2}}{g^{2}n}\,.
\end{equation}
Here $\rho$ is the cylindrical radius (the distance from the vortex core).
When $S=\pi/2$, the velocity matrix $\hat{v}$ is diagonal and the
probe fields are decoupled.

\subsection{Creation of the second probe field with optical vortex}

Since the group velocity matrix $\hat{v}$ is not necessarily diagonal,
the individual probe fields $\mathcal{E}_{1}$ and $\mathcal{E}_{2}$
generally do not have a definite group velocity. Only their combinations
$\chi^{\pm}$ for which $\hat{v}\chi^{\pm}=v^{\pm}\chi^{\pm}$ propagate
with the definite velocities $v^{\pm}$ in the atomic cloud. If $v^{+}=v^{-}$,
this leads to the mixing between probe fields $\mathcal{E}_{1}$ and
$\mathcal{E}_{2}$.

Suppose that a single component of the monochromatic probe beam $\mathcal{E}_{1}=
\mathcal{E}_{1}(0,\rho,\varphi,t)\sim e^{-i\Delta\omega t}$
is incident on the atomic cloud at $z=0$, with $\Delta\omega$ being the deviation of the frequency of
the probe field $\mathcal{E}_1$ from its central frequency $\omega_1$. We use cylindrical coordinates
with cylindrical radius $\rho$, azimuth  $\varphi$ and longitudinal position $z$.
The atomic gas is considered to be
uniform along the propagation direction $z$ form the entry point of the probe beam at $z=0$ to its exit at $z=L$.
At the end of the cloud (at $z=L$) the transmitted fields are
$\mathcal{E}_{1}(L,\rho,\varphi,t)=T_{1}(\rho,\varphi)\mathcal{E}_{1}(0,\rho,t)$
and $\mathcal{E}_{2}(L,\rho,\varphi,t)=T_{2}(\rho,\varphi)\mathcal{E}_{1}(0,\rho,t)$,
where $T_{1}(\rho,\varphi)$ and $T_{2}(\rho,\varphi)$ are the corresponding
transmission amplitudes. Equation (\ref{eq:result-2}) can be written
in the following form for the monochromatic probe fields and the control
beams given by Eqs.~(\ref{eq:Rabi-1})--(\ref{eq:Rabi-2}):
\begin{equation}
\partial_{z}\mathcal{E}=i(K_{0}+K_{x}\sigma_{x}+K_{y}\sigma_{y})\mathcal{E}\,,\label{eq:sigma}
\end{equation}
where $\sigma_{x}$ and $\sigma_{y}$ are Pauli matrices and
\begin{eqnarray}
K_{0} & = & \frac{\Delta\omega}{c}+\frac{\Delta\omega}{2}\left[\frac{1}{v^{+}}+\frac{1}{v^{-}}+i\frac{2L}{\alpha}\Delta\omega\left(\frac{1}{(v^{+})^{2}}+\frac{1}{(v^{-})^{2}}\right)\right]\label{eq:k0}\\
K_{x} & = & \cos(S+l\varphi)\frac{\Delta\omega}{2}\left[\frac{1}{v^{+}}-\frac{1}{v^{-}}+i\frac{2L}{\alpha}\Delta\omega\left(\frac{1}{(v^{+})^{2}}-\frac{1}{(v^{-})^{2}}\right)\right]\\
K_{y} & = & -\sin(S+l\varphi)\frac{\Delta\omega}{2}\left[\frac{1}{v^{+}}-\frac{1}{v^{-}}+i\frac{2L}{\alpha}\Delta\omega\left(\frac{1}{(v^{+})^{2}}-\frac{1}{(v^{-})^{2}}\right)\right]\,.\label{eq:ky}
\end{eqnarray}
Here the losses enter via the optical density
\begin{equation}
\alpha=2\frac{g^{2}nL}{c\gamma}\,.
\end{equation}
Equation.~(\ref{eq:sigma}) has the plane wave solutions $\mathcal{E}\sim e^{i\Delta kz}$
with
\begin{equation}
\Delta k=K_{0}\pm K_{\bot}\,,
\end{equation}
where
\begin{equation}
K_{\bot}=\sqrt{K_{x}^{2}+K_{y}^{2}}\,.
\end{equation}
The spatial development of monochromatic probe fields is described
by Eq.~(\ref{eq:sigma}) providing a formal solution $\mathcal{E}(z)=e^{i(K_{0}+K_{x}\sigma_{x}+K_{y}\sigma_{y})z}\mathcal{E}(0)$.
Thus one can relate the two-component probe field at the entrance
and exit points as
\begin{equation}
\mathcal{E}(L)=e^{iK_{0}L}\left[\cos(K_{\bot}L)+i\frac{K_{x}\sigma_{x}+K_{y}\sigma_{y}}{K_{\bot}}\sin(K_{\bot}L)\right]\mathcal{E}(0)\,.
\end{equation}
Since the two-component probe field at the entrance ($z=0$) is $\mathcal{E}(0)=(1,0)^T$ ,
the transmission amplitudes read
\begin{eqnarray}
T_{1} & = & e^{iK_{0}L}\cos(K_{\bot}L)\label{eq:t1}\\
T_{2} & = & i\frac{K_{x}+iK_{y}}{K_{\bot}}e^{iK_{0}L}\sin(K_{\bot}L)\,.\label{eq:t2}
\end{eqnarray}

Let us choose the control fields $\Omega_{11}$ and $\Omega_{22}$
to be the first-order Laguerre-Gaussian beam with $l=1$, the other
two control fields beams being the plane waves, $|\Omega_{12}|=|\Omega_{21}|=\mathrm{const}$.
In this case one has 
\begin{equation}
\frac{|\Omega_{11}|}{|\Omega_{12}|}=\frac{|\Omega_{22}|}{|\Omega_{21}|}=a\frac{\rho}{\sigma}e^{-\rho^{2}/\sigma^{2}}
\end{equation}
where $\rho$ is the cylindrical radius (the distance from the vortex
core), $\sigma$ represents the beam width and $a$ defines the relative
strength of the vortex and non-vortex beams. Expanding $T_{2}$, the
first term in the power series of $\rho$ reads
\begin{equation}
T_{2}(\rho,\varphi)\approx-2i\frac{\Delta\omega L}{v_{0}(0)}a\frac{\rho}{\sigma}\cos(S)\exp\left[-il\varphi-iS+i\frac{\Delta\omega L}{c}+i\frac{\Delta\omega L}{v_{0}(0)}\left(1+i\frac{2}{\alpha}\frac{\Delta\omega L}{v_{0}(0)}\right)\right]\left(1+i\frac{4}{\alpha}\frac{\Delta\omega L}{v_{0}(0)}\right)\label{eq:t2-approx}
\end{equation}
Equation (\ref{eq:t2-approx}) shows that the transmission amplitude
$T_{2}(\rho,\varphi)$ increases linearly with the distance $\rho$
and contains a vortex phase factor $-l\varphi$. Thus in a vicinity
of the vortex vore the generated second beam looks very much like
the Laguerre-Gause beam.

\begin{figure}
\includegraphics[width=0.6\textwidth]{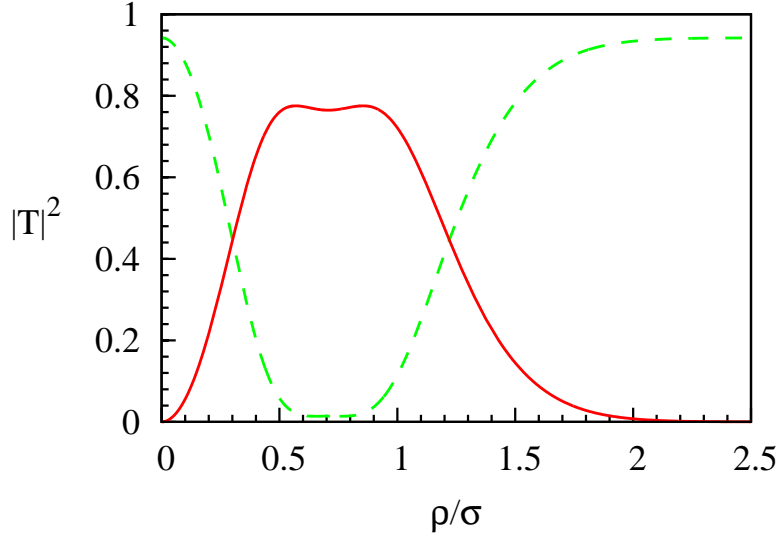}
\caption{(Color online) Dependence of the transmission probabilities
  $\left|T_{1}\right|^{2}$ (dashed green) and $\left|T_{2}\right|^{2}$ (solid
  red) on the dimensionless distance from the vortex core $\rho/\sigma$.
  Transmission probabilities are calculated using
  Eqs.~(\ref{eq:t1})--(\ref{eq:t2}) for the phase $S=0$, the parameter $a=1$
  and the optical density $\alpha=100$. The detuning frequency $\Delta\omega$
  is choosen such that the equality $\Delta\omega
  L(1/v^{-}(\rho)-1/v^{+}(\rho))=\pi$ holds at the radius
  $\rho=\sigma/\sqrt{2}$ where the difference between two eigenvalues of group
  velocity is maximum. This gives $\Delta\omega L/v_{0}(0)\approx1.22$.  We
  seek to maximize the difference between eigenvalues because larger difference
  leads to more effective creation of the second probe beam.}
\label{fig: trans}
\end{figure}

In the whole range of distances $\rho$ the transmission probabilities
are shown in Fig.~\ref{fig: trans} for the phase $S=0$ . The non-adiabatic
losses are seen to decrease the maximum amplitude of the second probe
beam. It is noteworthy that the detuning frequency $\Delta\omega$
and the length $L$ enter the transmission probabilities only in the
combination $\Delta\omega L/v_{0}(0)$. Thus increasing the sample
length has the same effect as increasing the detuning. 

\subsection{Estimation of the maximum detuning}

As can be seen from the last term in Eq.~(\ref{eq:result-2}), the detuning $\Delta\omega$ introduces
a finite life time for the polariton. The last term in Eq.~(\ref{eq:result-2}) yields the decay rate inversely proportional
to the group velocity, therefore, the life time of the polariton is determined by the minimum of the group velocity. However, the group velocity cannot be arbitrarily small because of the adiabaticity requirement.
Thus we assume the minimum group velocity to be of the order
of $v_{0}(\rho)$ at $\rho=0$, i.e. $v_{\mathrm{min}}\sim v_{0}(0)$.
The life-time of the polariton is then
\begin{equation}
\tau^{-1}=\gamma(\Delta\omega/\Omega(0))^{2}
\end{equation}
The requirement that the group velocity should be not too small constrains
the parameters of the beams: If $a>\sqrt{2e}$ and $S=0$ or $S=\pi$,
the minimum group velocity is $0$.

We can assume that the characteristic time of polariton evolution
is the time required to cross the atomic cloud of the length $L$.
In the case of copropagating control and probe beams the characteristic
time is $\tau_{\mathrm{pol}}=L/v_{\mathrm{min}}\sim L/v_{0}(0)$.
The characteristic time of the polariton evolution $\tau_{\mathrm{pol}}$
should be much smaller than the polariton life time $\tau$. From
this condition we obtain a constraint on the detuning 

\begin{equation}
\Delta\omega\ll\frac{\Omega(0)}{\sqrt{\gamma\tau_{\mathrm{pol}}}}=\Omega(0)\sqrt{\frac{v_{0}(0)}{\gamma L}}\,.\label{eq:cond-delta}
\end{equation}
Equation (\ref{eq:cond-delta}) can be written in the form
\begin{equation}
\frac{2}{\alpha}\left(\frac{\Delta\omega L}{v_{0}(0)}\right)^{2}\ll1\,.
\end{equation}
This condition also follows from the requirement that the last term
in Eqs.~(\ref{eq:k0})--(\ref{eq:ky}) should be small compared to
other terms.

On the other hand, the optical density $\alpha$ of the atomic cloud
is constrained from below. The maximum amplitude of the probe field
$\mathcal{E}_{2}$ is when $\Delta\omega L(1/v^{-}-1/v^{+})\sim\pi$.
Assuming that $(1/v_{\mathrm{gr}}^{-}-1/v_{\mathrm{gr}}^{+})\sim1/v_{0}(0)$
we get $\Delta\omega\sim\pi v_{0}(0)/L$. Substituting into Eq.~(\ref{eq:cond-delta})
we obtain
\begin{equation}
L\gg\frac{\pi^{2}\gamma v_{0}(0)}{\Omega^{2}(0)}\,.
\end{equation}
This condition means that the optical density must be sufficiently
large: $\alpha\gg2\pi^{2}\approx20\,.$

Note, that non-resonant transitions to other hyperfine levels of the electronic
excited state can become important when the hyperfine splitting is not large
enough compared with the detuning $\Delta\omega$, with the Rabi frequencies of
control fields or with the natural decay rate. Influence of the non-resonant
transitions has been studied in Refs.~\cite{Sheremet2010,Deng2002} showing that
the position of EIT resonance is shifted and the medium is no longer perfectly
transparent.  Thus, influence of the non-resonant transitions makes the losses
larger than those shown in Fig.~\ref{fig: trans}.

\section{\label{sec:conclusions}Concluding remarks}

We have analyzed the manipulation of slow light with the OAM by using
control laser beams with and without optical vortices. We have considered
a situation where the atom-light interaction represents a double tripod
scheme involving four control laser beams of different frequencies
which renders medium transparent for a pair of low intensity probe
fields. Under the conditions of electromagnetically induced transparency
(EIT) the medium support a lossless propagation of slow light quasiparticles
known as dark state polaritons. In the case of double tripod setup
these polaritons become two-component (spinor) quasiparticles, involving
both probe fields. By properly choosing the control lasers one can
generate a tunable coupling between the constituent proble fields.
Here we have studied the interaction between the probe fields when
two control beams carry optical vortices of opposite helicity. As
a result, a transfer of the optical vortex from the control to the
probe fields takes place. Notably, the transfer of the optical vortex
occurs during the polariton propagation without switching off the
control beams. This feature is missing in a single tripod scheme where
the optical vortex can be transferred from the control to the probe
field only during either the storage or retrieval of light. The manipulation
of spinor slow light with the optical vortices has potential application
in the optical information processing in quantum atomic gases.

\begin{acknowledgments}
This work has been supported by the project TAP LLT 01/2012 of the Research
Council of Lithuania, the National Science Council of Taiwan and the EU FP7
IRSES project COLIMA (contract PIRSES-GA-2009-247475). IAY acknowledges a
support by the TLL Grant No. NSC 102-2923-M-007-001 of National Science Council
of Taiwan.
\end{acknowledgments}

\end{document}